\documentstyle[epsfig]{mn}
\begin{document}
\LARGE
\normalsize
\title{The infrared counterpart of the Z source GX~5--1}

\author[P.~G. Jonker \& R.~P. Fender \& N.C. Hambly \& M. van der Klis]
{P. G. Jonker$^1$\thanks{email : peterj@astro.uva.nl}
R. P. Fender$^1$\thanks{email : rpf@astro.uva.nl}
N. C. Hambly$^2$\thanks{email : nch@roe.ac.uk}
M. van der Klis$^1$\thanks{email : michiel@astro.uva.nl}\\
$^1$ Astronomical Institute ``Anton Pannekoek'',
University of Amsterdam, and Center for
High-Energy Astrophysics, Kruislaan 403,\\ 1098 SJ, Amsterdam, The
Netherlands\\
$^2$ Institute for Astronomy, University of Edinburgh, Royal
Observatory, Blackford Hill, Edinburgh, EH9 3HJ, Scotland, UK\\
}
\maketitle

\begin{abstract}

We have obtained UKIRT infrared observations of the field of the
bright Galactic Z source GX~5--1. From an astrometric plate solution
tied to Tycho-ACT standards we have obtained accurate positions for
the stars in our field which, combined with an accurate radio
position, have allowed us to identify the probable infrared
counterpart of GX~5--1. Narrow--band photometry marginally suggests
excess Br$\gamma$ emission in the counterpart, supporting its
association with an accretion--disc source.  No significant
variability is observed in a limited number of observations.  We
compare the H and K magnitudes with those of other Z sources, and
briefly discuss possible sources of infrared emission in these
systems.

\end{abstract}


\section{Introduction}
GX~5--1 is the second brightest persistent Galactic X-ray source. The
source is well studied in X-rays; it was classified as a Z source on
the basis of the pattern it traces in a colour-colour diagram and its
timing properties (Hasinger \& van der Klis 1989). Quasi-periodic
oscillations with frequencies of 13--50 Hz, 6 Hz, and 200--800 Hz were
detected in the X-ray lightcurves (van der Klis 1985a,b; Lewin et
al. 1992; Wijnands et al. 1999, respectively).  Naylor, Charles, \&
Longmore (1991) identified several candidate infrared counterparts of
GX~5--1. Since it is located near the Galactic centre, source
confusion and heavy optical obscuration hinder the classification.

GX~5--1 is also a radio source (Braes, Miley, \& Schoenmaker 1972;
Grindley \& Seaquist 1986; Penninx et al. 1988; Berendsen et
al. 2000), like all Z sources (Hjellming \& Han 1995; Fender \& Hendry
2000). The radio emission is likely to arise in a compact jet from the
system. The radio counterpart allows for extremely accurate position
measurements.

Study of the Z sources has been hampered in most cases by the lack of
reliable optical and/or infrared counterparts. For example, Deutsch et
al. (1999) showed that the proposed infrared counterpart (Tarenghi \&
Reina 1972) of another persistently X-ray bright Z source, GX~17+2 was
not consistent with the position of its radio
counterpart. Furthermore, they detected a faint star close to the
proposed counterpart. Callanan, Fillipenko, \& Garcia (1999) reported
variability of about 3.5-4 mag in the K band for the latter, providing
additional evidence for its classification as the counterpart.

In this Letter, we present United Kingdom Infrared Telescope (UKIRT)
infrared (IR) observations of the X-ray source GX~5--1. We resolve the
previously reported counterparts, and show that the radio position is
consistent with only one of them.

\begin{figure*}
\leavevmode\epsfig{file=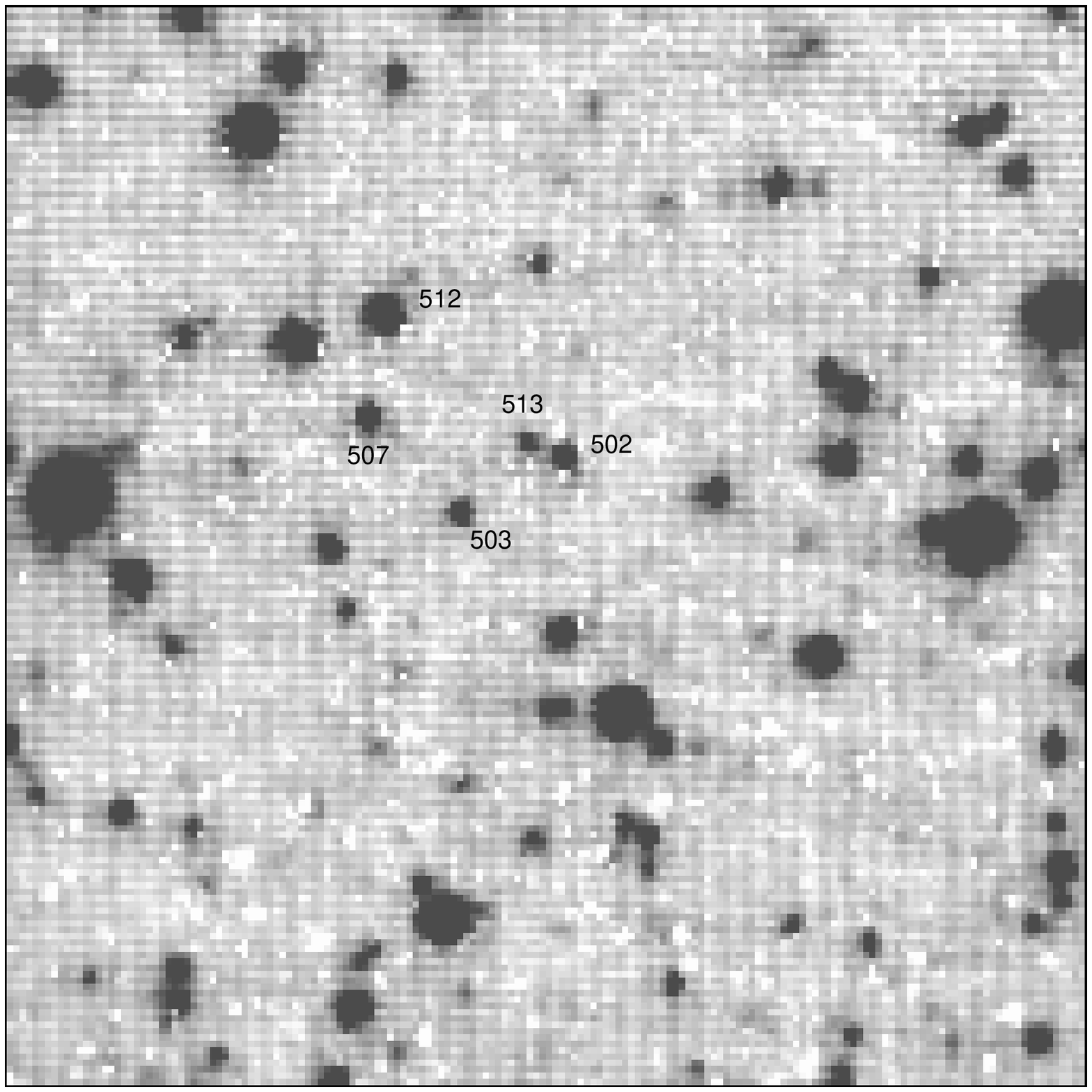,width=6cm}
\quad
\leavevmode\epsfig{file=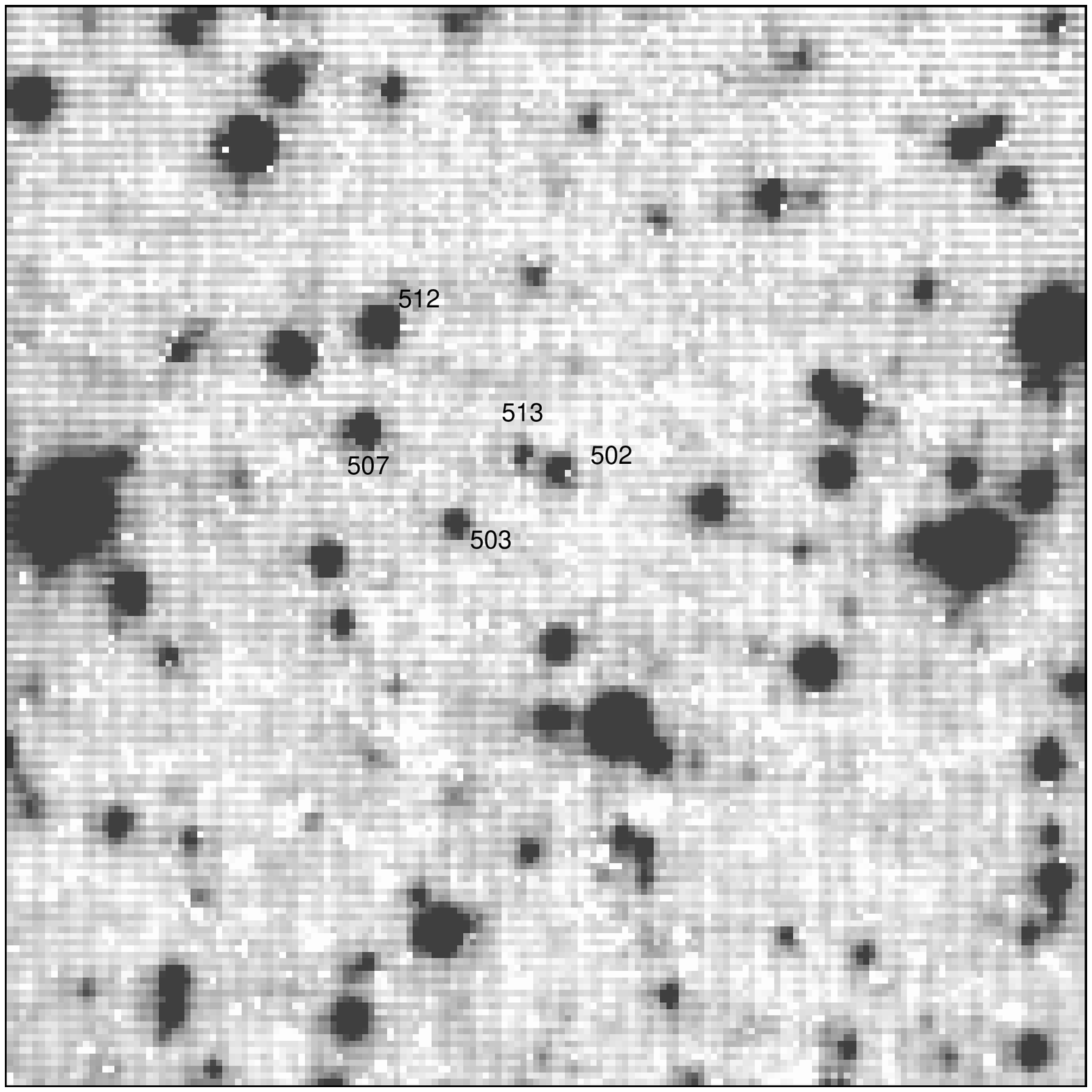,width=6cm}
\caption{Combined images obtained in the K (left) and H (right) bands on May
23, 1999 showing the field of GX~5--1.
The labels of the stars are those of Naylor et
al. (1991).}
\label{finding_chart}
\end{figure*}

\section{Observations, analysis and results}

We observed the field of GX~5--1 with UKIRT. Observations were taken
in H, K, and in a narrow filter around the Brackett Gamma line
(Br$\gamma$) in 1999 May and October. A log of the observations can be
found in Table~\ref{log}. The observations of 1999, May 23 were
obtained using the IRCAM3 camera; the frames consist of 256 $\times$
256 pixels with a pixel size of 0.286 arc seconds. The observations of
1999, October 8 and 13 were performed using the UFTI camera; the UFTI
frames consist of 1024 $\times$ 1024 pixels, with a pixel size of
0.0909 arc seconds. The Br$\gamma$ narrow filter is centred on the
wavelength of the Br$\gamma$ line (2.166 micron; 50\% of the light was
obtained in the wavelength range 2.151--2.171 $\mu$m in case of IRCAM3
observations, and in the range 2.155--2.177 $\mu$m in case of UFTI
observations). The night was photometric only during the 1999, May 23
observations. The exposure time used in the Br$\gamma$ filter was 100
seconds, and in the H and K filter band a 10 second exposure was
used. On 1999, October 13 an observation time of 100 seconds was used
in the K band.

\begin{table}
\caption{A log of our UKIRT observations of GX~5--1.}
\label{log}
\begin{tabular}{cccc}
Date (1999) & MJD & Filter & No. of exposures \\

May 23 & 51321 & K, H, Br$\gamma$ & 5, 5, 10 \\
Oct 8 & 51459 & Br$\gamma$ & 9\\
Oct 13 & 51464 & K & 6\\

\end{tabular}
\end{table}

\subsection{Photometry}
All images were dark subtracted. Five (or three in case of the 1999,
October 13 observations) dithered IR frames were used to calculate a
sky image. This dark subtracted sky image was subtracted from the dark
subtracted image after scaling it to the object image level. The
resulting images were flatfielded, where the flatfield image was
obtained by normalizing the combined five (or three) dithered
images. The reduced images were aligned and combined.

In Fig.~\ref{finding_chart}, we show the observed field in both the K
and H filter bands. Clearly visible is that objects 502 and 513, which
were blended in the images of Naylor et al. (1991), are resolved into
two separate stars.

We used three reference stars in the field (503, 507, 512 in the
images of Naylor et al. 1991, see Fig.~\ref{finding_chart}) to obtain
differential magnitudes of the counterpart. These stars were
calibrated by observing the standard star HD~161903 on 1999, May 23.
To obtain the differential magnitudes, we used the point spread
fitting (psf) routine to both the reference and object stars using the
DoPHOT package (Schechter, Mateo, \& Saha 1993). The magnitudes
derived in this way did not differ significantly from the magnitudes
derived using aperture photometry. We corrected the magnitudes for the
airmass dependent atmospheric extinction in the H and K band. The
magnitudes we derived in the H and K bands are listed in
Table~\ref{mags}. We seached for variability on timescales of $\la10$
minutes and in between the observations, but no significant
variability was observed in any of the stars listed in
Table~\ref{mags}. We determined an upper limit on photometric
variability on timescales of $\la10$ minutes of 0.6 magnitudes in the
H and K band.  In Table~\ref{mags} we also list the flux densities
obtained in the H and K filter bands, as well as the Br$\gamma$--K
instrumental magnitude difference.  No standard magnitude in the
Br$\gamma$ filter is known for the star HD~161903, therefore we could
only calculate instrumental magnitudes in this band.

Accretion disks are known to sometimes produce Br$\gamma$ emission
lines (eg. Bandyopadhyay et al. 1997;1999). Therefore, if a strong
Br$\gamma$ emission line is present in the accretion disk of GX~5--1
the counterpart could appear brighter in this filter.  We compared the
instrumental Br$\gamma$--K colour of the stars 502, 513, and our
reference stars (503, 507, and 512); these are also listed in Table 2.
Star 513 seems to have a smaller instrumental Br$\gamma$--K colour,
although the effect is only marginally detected.  We also checked for
variability in the Br$\gamma$ band, but no significant variability was
found on timescales of minutes with an upper limit of 0.45 magnitude.

\begin{table*}
\caption{The observed magnitudes, flux densities in the H and K band,
and the H $-$ K colour on May 23, 1998. Additionally the Br$\gamma$
$-$ K instrumental magnitude colour is given.}
\label{mags}
\begin{tabular}{ccccccc}
Stars & H magnitude & Flux density H (mJy) & K magnitude & Flux
density K (mJy) & H--K & Br$\gamma$--K\\
502 & $13.3\pm0.1$& 4.9 & $12.6\pm0.1$& 6.0 & 0.7 & $0.74\pm0.08$\\
503 & $13.5\pm0.1$& 4.1 & $12.9\pm0.1$& 4.5 & 0.6 & $0.73\pm0.09$ \\
507 & $12.42\pm0.06$& 11.0 & $12.42\pm0.08$& 7.1 & 0.0 & $0.98\pm0.09$ \\
512 & $11.70\pm0.03$& 21.3 & $11.28\pm0.04$& 20.2 & 0.42 & $0.79\pm0.04$\\
513 & $14.1\pm0.2$& 2.3 & $13.7\pm0.2$& 2.2 & 0.4 & $0.6\pm0.1$\\
\end{tabular}
\end{table*}

\subsection{Astrometry}
We used the higher resolution UFTI images obtained on 1999, October 8
for our astrometry.  To define astrometric solutions for the IR frames
we used secondary astrometric standards derived from United Kingdom
Schmidt photographic plate material measured using the precision
microdensitometer SuperCOSMOS (eg. Hambly et al. 1998). The global
astrometric solution for the Schmidt plate was derived using the
Tycho--ACT reference catalogue (Urban, Corbin, \& Wycoff 1998), and
includes correction for non-linear systematic effects caused by the
mechanical deformation of the plates during exposure (eg. Irwin et
al. 1998). We used the "short red" survey plate R5803 (epoch 1979.5,
field number 521). These short exposures, taken at low galactic
latitudes, are far less crowded than the sky limited survey plates and
reach R$\sim$20 (as opposed to R$\sim22$ for the deep survey
plates). They are ideal for accurate astrometry of secondary standards
as faint as R=20 which overlaps with unsaturated objects on the IR
frames. The rms residual per ACT star in the global astrometric
photographic plate solution was $\sim0.2$ arcsec in both
coordinates. A solid--body linear plate solution (ie. 4--coefficient)
was derived between 7 stars in common between the photographic and IR
data, yielding a plate scale of 0.0903 arcsec/pix and rms errors per
secondary standard of $\sim0.1$ arcsec in either coordinate.  We
estimate that there will be no systematic zero--point errors in the
global IR array astrometric solution larger than $\sim0.25$ arcsec.
 
\begin{figure}
\centering
\leavevmode\epsfig{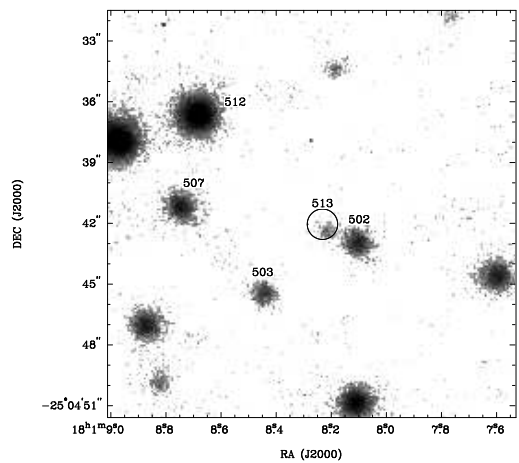}
\caption{Logarithmically scaled section of the combined Br$\gamma$
images obtained on 1999, October 8. The 3$\sigma$ error in the
astrometric solution is shown as a circle centred on the radio
position of GX~5--1. The labels of the stars are those of Naylor et
al. (1991).}
\label{5-1}
\end{figure}

Since the uncertainty in the radio position is small ($<$40 mas,
Berendsen et al. 2000) compared to the estimated uncertainty in the
astrometric solution, the overall uncertainty in the radio--infrared
alignment was estimated to be 0.25 arcsec. The coordinates of the
stars 502 and 513 are listed in Table~\ref{positions}. Comparing these
positions with the accurate radio position of GX~5--1 given by
Berendsen et al. (2000), we conclude that of the detected stars, star
513 is the only plausible counterpart of GX~5--1 (see Fig.~\ref{5-1}).

\begin{table}
\caption{Positions of the stars 502 and 513 obtained from a global
astrometric plate solution. The accurate radio position from Berendsen
et al. (2000) is also listed. In the last collumn the separation (d)
between the radio position and the position of star 502 and 513 is given.}
\label{positions}
\begin{tabular}{ccccc}
Stars & RA & DEC & $\sigma$ & d \\
502 & 18:01:08.109 & -25:04:43.02 & 0.25''& $\sim1.9$''\\
513 & 18:01:08.222 & -25:04:42.46 & 0.25''& $\sim0.2$''\\
Radio & 18:01:08:233 & -25:04:42.044 & 0.04''\\
\end{tabular}
\footnotesize
\normalsize
\end{table}

\begin{table*}
\caption{K band magnitudes, distance estimates, estimates of
$\rm{N_H}$, calculated absolute $\rm{M_K}$, best estimates of the
$\rm{P_{orb}}$ and spectral type of different Z sources and GX~13+1.}
\label{zsources}
\begin{tabular}{ccccccc}
Stars & K band & D (kpc)$^1$ & $\rm{N_H}$ ($10^{21}\rm{cm^{-2}}$)$^1$&
$\rm{M_K}$ & $\rm{P_{orb}}$ (hr) & Companion Type\\ 

Sco~X--1 & 11.9 $^2$  & 2.8 & 2.9 & -0.5 & 18.9$^9$ & $<$G5 III$^7$
($\rm{M_K} \geq -1.4$)\\
GX~17+2  & 14.5 $^3$  & 7.5 & 17.3 & -1.1 & -- & -- \\
         & 18.5       &     &      & +2.9  & &\\
Cyg~X--2 & 13.8 $^2$  & 8.0 & 2.8 & -0.9  & 236.2$^8$& A9 III$^8$($\rm{M_K} \sim -0.7$) \\
GX~5--1  & 13.7       & 9.0 & 25.4 & -2.8  & -- & -- \\
GX~340+0 & 17.3 $^4$  & 11.0 & 50 & -1.0  & -- & -- \\
GX~349+2 &$\sim$14 $^5$& 5.0 & 8.8 & 0.0 & $\sim$22$^{5,10}$ or 14 d$^{11}$ & -- \\

GX~13+1  &$\sim$12 $^6$ & 7.0 & 25.4 & -3.8  & -- & K5 III$^7$($\rm{M_K} \sim -3.8$)\\

\end{tabular}
\footnotesize
\newline
1.\,Christian \& Swank (1997), Sco~X--1 from Bradshaw et al. (1999);
2.\,Priv. comm. T. Shabhaz;
3.\,Callanan et al. (1999), magnitude when flaring;
4.\,Miller et al. (1993);
5.\,Wachter \& Margon (1996);
6.\,Charles \& Naylor (1992);
7.\,Bandyopadhyay et al. (1999);
8.\,Casares et al.  (1998);
9.\,Gottlieb et al. (1975);
10.\,Barziv et al. (1997);
11.\,Southwell et al. (1996)
\end{table*}

\section{Discussion}
We have shown that star 513 of Naylor et al. (1991) is most likely the
IR counterpart of the low-mass X-ray binary (LMXB) GX~5--1, since its
position coincides with the accurate position of the radio
counterpart. Furthermore, the Br$\gamma$--K instrumental colour is
smaller for star 513, when compared to the reference stars in the
field. This effect, although only marginally detected might be caused
by the presence of a Br$\gamma$ emission line in the spectrum of star
513. The presence of an absorption line could result in a higher
Br$\gamma$--K colour, which might explain the higher value obtained
for star 507.

The H and K magnitudes we derived are significantly lower than the H
and K magnitudes derived by Naylor et al. (1991), but as they mention
in their paper they estimate systematic errors to play a role
(although these errors were estimated to be smaller than the
discrepancy with our results). Bandyopadhyay et al. (1999) obtained IR
spectra of the stars 502 and 503. Since they did not find evidence for
emission typical of an accretion disk in these two stars, they
suggested that star 513 might be the counterpart of GX~5--1.

The counterpart (star 513) did not vary significantly in any of the
three filters (H, K, and Br$\gamma$) we used. Its reddening
uncorrected H--K colour index is 0.4. Using the conversion of $N_H =
0.179 A_V 10^{22} cm^{-2}$ (Predehl \& Schmitt 1995) and the estimate
of $N_H$ for GX~5--1 of $\sim2.5\times10^{22} cm^{-2}$ (Christian \& Swank
1997; see Table ~\ref{zsources}) we obtain $A_V \sim 14$. Using the
relations found by Rieke \& Lebofsky (1985) we obtained an intrinsic
$(\rm{H}-\rm{K})_{\circ} = -0.5$.  This is bluer than stellar
(Tokunaga 2000), which may indicate an overestimate of the
extinction. Limiting the intrinsic emission in the near-infrared to be
no steeper than the Rayleigh-Jeans tail of a black-body implies $A_V
\la 12$.

In Table ~\ref{zsources} we compare the K-band absolute magnitudes of
the six Z sources plus the `hybrid Z/atoll' source GX~13+1, based on
the estimated distance and ${\rm N_H}$ to each source. There is a
rather large range, from as bright as $-3.8$ for GX~13+1 to possibly
as faint as $+2.9$ for GX~17+2 if the observed K magnitude in
quiescence is as faint as $\sim 18.5$ (Callanan et
al. 1999). Uncertainties in distance estimates and reddening are
likely to be significant at a level of about $\pm 1$ magnitude, and so
cannot account for the broad range. Several different components may
contribute significantly to the emission in the near--infrared; as a
guide to their significance (see below) we have also listed binary
orbital periods and, where available, the spectral types of the mass
donors in Table ~\ref{zsources}.  Thermal emission will be produced
both by the stellar companion and the accretion disc (for a discussion
of their relative contributions see also Bandyopadhyay et al. 1997;
1999). We may expect the accretion-disc contribution to depend on the
size of the disc (van Paradijs \& McClintock 1994), which in turn
should be a function of the orbital period of the system.  We note
that for the three systems with some attempt at spectral
classification of the mass donor there is a good agreement between the
absolute K band magnitudes derived and those expected for the
companion spectral class. This implies that GX~5--1 should contain a
relatively bright mass donor.  Luminosity class III was found for the
companion star in Sco~X--1 and Cyg~X--2 (see Table ~\ref{zsources}, and
references therein). Following the conjecture made by Hasinger \& van
der Klis (1989) that all Z sources have evolved companions, we assume
also luminosity class III for GX~5--1. The companion star in GX~5--1
is then most likely of spectral type K.

There may also be an additional contribution from infrared synchrotron
emission, as found in the black hole system GRS~1915+105 (Fender \&
Pooley 1998 and references therein). If at all, this should only occur
when the source is radio--bright. The Z sources are brightest at radio
wavelengths when they are observed on the Horizontal Branch (HB) in
the X-ray colour-colour diagram (Penninx et al. 1988; Hjellming \& Han
1995). Radio flaring in Z sources typically has amplitudes of a few
mJy (Hjellming \& Han 1995 and references therein); if the synchrotron
emission has a flat spectrum to the near-infrared we might observe a
(reddened) contribution of $\ga 1$ mJy at times. For GX~5--1 this
could cause up to 1 magnitude variability.

If star 513 is not the counterpart of GX~5--1, the counterpart must
have been $\ga$2.5 magnitudes fainter in the K band at the time of our
observations. Future spectroscopic observations and/or the detection
of variability should confirm star 513 as the counterpart.

To conclude, we have most likely identified the IR counterpart of the
bright Z-type X-ray source GX~5--1 based upon positional coincidence
with the radio counterpart, an identification which is supported by
marginal evidence for excess Br$\gamma$ emission. We have discussed
the possible origins of IR emission in this system and in the other Z
sources (and GX~13+1), and suggest that GX~5--1 may contain a KIII
mass donor.

\section*{Acknowledgments}
The UKIRT is operated by the Joint Astronomy Centre on behalf of the
U.K. Particle Physics and Astronomy Research Council. The data
reported here were obtained as part of the UKIRT Service Programme. We
would like to thank Sandy Leggett for the May 23 observations and
helpful comments in reducing the observations, John Davies for the
October observations, Paul Vreeswijk for help with the reduction of
the images, Tim Naylor for providing us with electronic versions
of the images presented in Naylor et al. (1991) facilitating
comparisons, and the referee, Phil Charles for his comments which
improved the paper.

\end{document}